\documentstyle[psfig]{mn}

\def\sp{{\sc spectrum}}

\def\met{[Fe/H]}

\def\deg{^{\circ}}

\def\gtsim{\:{_>\atop{^\sim}}\:}
\def\ltsim{\:{_<\atop{^\sim}}\:}

\def\teff{$T_{\rm{eff}}$}
\def\logg{$\log g$}
\def\met{[M/H]}
\def\vmicro{$V_{\rm{micro}}$}

\def\kms{\,km\,s$^{-1}$}

\def\1pos{$x_1,y_1$}
\def\2pos{$x_2,y_2$}
\def\3pos{$x_3,y_3$}
\def\4pos{$x_4,y_4$}

\def\sig68{$\sigma_{\rm{68}}$}

\def\td{$D_s(T_{\rm{eff}})$}
\def\mtf{$\tilde{T}_{\rm{eff}}$}

\title[Physical Stellar Parameters from Neural Networks]
{Physical Parameterization of Stellar Spectra: The Neural Network Approach}
\author[C.A.L.\ Bailer-Jones et al.]
{Coryn A.L.\ Bailer-Jones$^1$\thanks{Present Address: 
Mullard Radio Astronomy Observatories,
Cavendish Laboratory, Madingley Road, Cambridge, 
CB3 0HE, UK}\thanks{email: calj@mrao.cam.ac.uk},
Mike Irwin$^2$, Gerard Gilmore$^1$ and Ted von Hippel$^3$ \\
$^1$ Institute of Astronomy, Madingley Road, Cambridge, CB3 0HA, UK\\
$^2$ Royal Greenwich Observatory, Madingley Road, Cambridge, CB3 0EZ, UK\\
$^3$ Department of Astronomy, University of Wisconsin, Madison, WI 53706, USA\\
}
\date{Submitted April 1997, Accepted July 1997}
\pubyear{1997}

\begin{document}

\maketitle
\begin{abstract}
We present a technique which employs artificial neural networks to
produce physical parameters for stellar spectra. A neural network is
trained on a set of synthetic optical stellar spectra to give physical
parameters (e.g.\ \teff, \logg, \met). The network is then used to
produce physical parameters for real, observed spectra.

Our neural networks are trained on a set of 155 synthetic spectra,
generated using the \sp\ program written by Gray (Gray \& Corbally
1994, Gray \& Arlt 1996)\nocite{gray_94b}\nocite{gray_96a}.  Once
trained, the neural network is used to yield \teff\ for over 5000
B--K spectra extracted from a set of photographic objective prism
plates (Bailer-Jones, Irwin \& von Hippel
1997a)\nocite{bailerjones_97a}.  Using the MK classifications for
these spectra assigned by Houk (1975, 1978, 1982,
1988)\nocite{houk_75a}\nocite{houk_78a}\nocite{houk_82a}\nocite{houk_88a},
we have produced a temperature calibration of the MK system based on
this set of 5000 spectra.  It is demonstrated through the metallicity
dependence of the derived temperature calibration that the neural
networks are sensitive to the metallicity signature in the real
spectra.  With further work it is likely that neural networks will be
able to yield reliable metallicity measurements for stellar spectra.
\end{abstract}

\begin{keywords}
methods: data analysis, numerical - stars: fundamental parameters
\end{keywords}

\section{Introduction}

The MK classification system was first proposed in its current form in
1943 by Morgan, Keenan \& Kellman (1943)\nocite{morgan_43a}, and has
since undergone a number of revisions (e.g.\ Keenan \& McNeil
(1976)\nocite{keenan_76a}, Morgan, Abt \& Tapscott
(1978)\nocite{Morgan_78a}).  MK classification is the only widely used
system for stellar spectral classification. Over its history it has
contributed towards a number of important developments in astronomy,
such as the further development of the now-famous HR diagram (Hertzprung 1911,
Russell 1914) and the identification of anomalous stars.  Currently, MK
classification is largely used as a tool in the preliminary analysis
of unusual stars, and in selecting stellar samples for further study.

An often-stated advantage of the MK system is that its
classifications, often based upon the visual inspection of spectra,
are static because they are based on a set of standards.  However, a
given spectrum may be classified differently by different people, and
any one person may also classify a given spectrum differently at
different times.  These problems of subjectivity could be partially
alleviated through the use of automated classifiers (von Hippel
et~al.\ 1994\nocite{vonhippel_94a}, Bailer-Jones et~al.\
1997a\nocite{bailerjones_97a}).  Automated classifiers could also
produce quantitative errors associated with their classifications.
Another problem with the MK system is that it lacks a well-defined
metallicity parameter, whereas metallicity variations are known to
have a significant effect on the appearance of high ($\sim$ 1\,\AA)
resolution spectra. This limits the system to classifications of
bright, nearby stars which show only limited metallicity variations.

Attempts to extend and revise MK classification (e.g.\ Corbally, Gray
\& Garrison 1994)\nocite{corbally_94a} may well prove valuable, but as
our understanding of stellar spectra grows, particularly from
computational work with model atmospheres and synthetic spectra, it
becomes increasingly desirable to obtain reliable physical
parameterizations of stars.  Any classification system is a compromise
between retaining the full information in the spectrum and the need
for a compact summary of it.  The optimal `summary' is of course given
by the physical parameters.  Advances in computational power and data
storage since the inception of MK classification mean that it is now
practicable to process and store large numbers of spectra. The
development of fast, automated classifiers will mean that it is
feasible to `classify' large numbers of stellar spectra in terms of
their physical parameters and to re-classify them rapidly whenever
stellar models are improved.  Physical parameters should be obtained
 from an original spectrum, rather than an empirical classification, as
the latter may well disregard certain spectral features which later
turn out to be important.  One of the advantages of the MK
classification system is that it is an empirical system based on
unchanging standards, whereas any direct parameterization of a
spectrum depends upon the quality of stellar models and will change as
these improve.  MK classifications could remain as useful labels
giving a rough `feel' for a spectrum, and the work of von Hippel
et~al.\ (1994)\nocite{vonhippel_94a} and Bailer-Jones et~al.\ 
(1997a)\nocite{bailerjones_97a} has shown that reliable automated MK
classification is possible.  In this paper we demonstrate how we
extend our automated techniques to the determination of physical
parameters directly from an observed spectrum.

\section{Observational Data}

The observational stellar spectra used in this project were taken from
objective prism plates obtained in the Michigan Spectral
Survey (Houk 1994)\nocite{houk_94a}. This was an objective prism survey of the
whole southern sky ($\delta < +12\deg $) from the Curtis Schmidt
telescope at the Cerro Tololo Interamerican Observatory in Chile.  We
were loaned 100 of these plates by Houk, which we digitized using the
APM facility in Cambridge. 
The output from the APM is an optical density.
Cawson et~al.\ (1987)\nocite{cawson_87a} showed that the optical density
calculated by the APM is, to a good approximation, linearly related to 
the incident intensity on the photographic plate for the typical range of 
optical density encountered for this type of extraction.
Once scanned, the
digitized plates were reduced and the
spectra extracted off-line. This yielded a set of over 5000 spectra
over the approximate spectral range 3800--5200\,\AA\ at a resolution
of $\sim$3\,\AA. This data set covers a wide range of spectral types
(B2--M7) for luminosity classes III, IV and V.  The Michigan Spectral
Survey was designed to be a re-classification of the Henry
Draper stars. In keeping with previous usage, we shall refer to our
set of 5000 spectra from this survey as the MHD (Michigan Henry
Draper) spectra.  Further details of this data set and the spectral
extraction technique are discussed in Bailer-Jones, Irwin \& von Hippel
(1997b)\nocite{bailerjones_97b}. These spectra have been used in a
related project to automate MK stellar classification using a neural
network (Bailer-Jones et~al.\ 1997a)\nocite{bailerjones_97a}.

\section{Neural Networks}

\begin{figure}
\centerline{
\psfig{figure=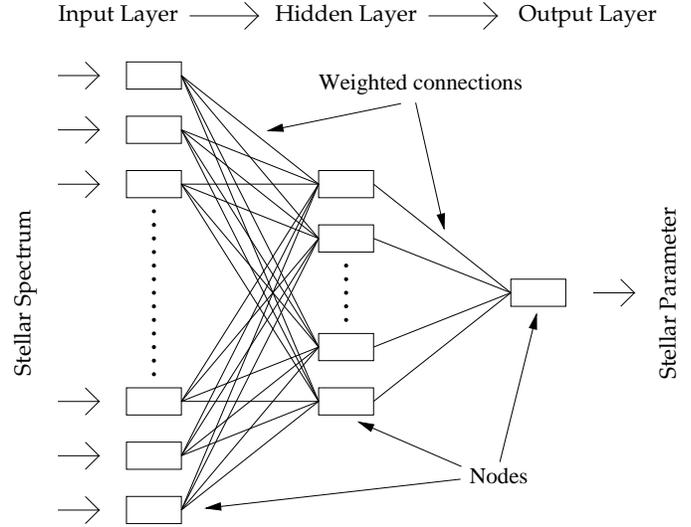,angle=270,width=0.5\textwidth}
}
\caption{Neural network architecture. A neural network consists of
layers of nodes linked by weighted connections (weights).  
Each node in the hidden
and output layers performs a non-linear weighted sum of its inputs which it
then passes to its output.
Therefore the final output from the neural network is a non-linear
function of the network inputs parameterized by the weights. 
The weights are found by `training' the neural network using
a set of inputs for which the ideal outputs (the `target outputs') are known.
This training process is therefore equivalent to interpolating
the training data to find the underlying function relating the inputs
to the output.
 }
\label{typnetarch.p3}
\end{figure}
A neural network is a software device which can be
trained to give a non-linear parameterized mapping between a number of
inputs (e.g.\ a complete stellar spectrum) and one or more outputs
(e.g.\ parameters such as \teff\ or MK spectral type).
Figure~\ref{typnetarch.p3} shows a neural network architecture with a
single output.  Each node in the input layer holds a value, $I_i$. In
our application the vector of inputs, $(I_1,I_2,\ldots,I_i,\ldots)$,
is a stellar spectrum and the output is the effective temperature, \teff.  
Each of the input nodes connects to every node
in the next layer of nodes, the `hidden' layer, and each of these
connections has a weight, $w_{i,j}$, associated with it.  The $j^{th}$
node in the hidden layer forms a weighted sum of its inputs,
given by $x_j = \sum_i w_{i,j}I_i$.  It then passes this sum through a
non-linear sigmoid transfer function to give the final output from
this node, $H_j = (1 + e^{-\lambda x_j})^{-1}$, where $\lambda$ is
some constant.  These outputs from the nodes in the hidden layer then
serve as the inputs to the node in the output layer, which performs
the same processing.  The output from this final node is the network
output.  The non-linearity of the sigmoid function means that the
network output is a non-linear function of the inputs. It can be
shown that neural networks can model functions of arbitrary complexity
(e.g.\ Bishop (1995) and references therein).  Neural networks are
therefore useful in a wide range of data modelling applications.

The key to producing the desired output from the network when it is
presented with a certain input is to set the network parameters -- the
weights -- to their correct values. This is known as `training' the
network and requires a set of input data for which the associated
ideal outputs (the `target outputs') are known.  Training takes place
as follows.  The weights are initially set with random values over a
small range. Thus when a spectrum is fed into the network, the output
will also be random. By comparing this output with what it
should be (the target output), we can adjust the weights to give an
output which is closer to the target value. This is repeated for each
input/output pair in the training data set.  The network is trained
iteratively by successive passes of the training data through the
network, and on each pass the weights are perturbed towards their
optimal values.  Specifically, the network training is performed by
minimizing the error function $E = \sum_k(O_k - T_k)^{2}$,
where $O_k$ is the network output for a given input spectrum and $T_k$
is the target output for that spectrum. Thus we can think of training
a neural network as an $N$-dimensional minimization problem in which
we want to find those values of the $N$ network weights which minimize
$E$. Once the weights have been found they are fixed and the neural
network can be applied to any number of 
new inputs for which the outputs are not known.
Training the neural network is simply the process of interpolating
the multi-dimensional training data in order to produce an input-output
mapping characteristic of the problem represented by these data.

The output from a neural network is some non-linear function of all of
the network inputs.  In our particular application this means that the
values of physical stellar parameters which the neural network gives
at its output are based on the appearance of the whole spectrum: we do
not have to tell the network in advance which spectral lines are
relevant.  Based only upon the training data, the neural network will
learn which wavelengths are more significant than others in
determining the correct spectral parameters, and will express this by
assigning appropriate values to the network weights.

Neural networks have been applied in a number of areas of astronomy,
as reviewed by Storrie-Lombardi \& Lahav
(1994)\nocite{storrie-lombardi_94b} and Miller
(1993)\nocite{miller_93a}.  Further theoretical details of neural
networks can be found in, for example, Hertz, Krogh \& Palmer
(1991)\nocite{hertz_91a}, Bishop (1995)\nocite{bishop_95a}, Lahav
et~al.\ (1996)\nocite{lahav_96a} and Bailer-Jones
(1996)\nocite{bailerjones_96a}.

\section{Calibration Procedure}

In the following sections, theoretical stellar spectra of given
physical parameters (such as \teff\ and \logg) are generated and
processed to the same `flux' and dispersion scales as the observed MHD
spectra described in the previous section. A neural network is trained
on these synthetic spectra with their physical parameters 
as the network target outputs.  Once trained, the neural network is
used to give physical parameters for the MHD stars. This provides a
means of rapidly and easily obtaining physical parameters for a large
number of spectra without calibration via the MK system.  However, a
statistical comparison of these physical parameters with the known MK
classifications of the MHD stars provides a calibration between the MK
system and physical stellar parameters.  Training the network is very
quick ($\sim10$ seconds on a {\sc SUN} Sparc 10), and applying it is
even faster, so it is not prohibitive to re-train the network as
improved models are obtained.

\section{Synthetic Spectra}

\subsection{Generation}\label{generation}

The synthetic spectra were created using the \sp\ program written by
Gray (Gray \& Corbally 1994,
Gray \& Arlt 1996)\nocite{gray_94b}\nocite{gray_96a}.
This program computes a synthetic spectrum under the assumption of
Local Thermodynamic Equilibrium (LTE) using a model stellar
atmosphere.  We used the fully blanketed models calculated by Kurucz
(1979, 1992)\nocite{kurucz_79a}\nocite{kurucz_92a}.  The model
atmosphere is a tabulation of temperature and pressure at a range of
`mass depths' in the stellar photosphere, calculated on the basis of a
variety of sources of opacity. These model 
atmospheres are characterized by four
parameters: metallicity, \met; microturbulence velocity, \vmicro;
surface gravity, \logg; effective temperature, \teff. Each spectrum is
uniquely labelled by these four parameters.  To calculate the spectrum
emergent from the model atmosphere, \sp\ also requires a table of
atomic and molecular species which lists their relative abundances,
masses, and ionization energies (or disassociation energies for
molecules). \sp\ is also given a line list, listing, for each line in
the spectrum, the atomic or ionic species producing the line, the
energy of the two bound electron energy levels involved in the
transition, the oscillator strength of the transition and a damping
factor. From these, \sp\ calculates the densities of electrons, atoms
and ions at different layers in the atmosphere, from which it
determines the atmospheric opacity as a function of optical depth and
wavelength using a number of different opacity sources (Rayleigh
scattering, electron scattering, bound-free opacities etc.). Under the
assumption of LTE, the source function, $S_{\nu}$, is equal to the
Plank function, and the latter can be computed directly from the
temperatures tabulated in the model atmosphere.  Using the source
function and opacities, \sp\ calculates the synthetic spectrum in
small wavelength portions. This is done by calculating the continuum
and line absorption strengths at each wavelength and evaluating a
Voigt broadening profile for the lines using van der Waals, natural,
and quadratic Stark broadening.  More details of \sp\ can be found in
Gray \& Corbally (1994)\nocite{gray_94b}.

\begin{table}
\begin{center}
\caption{Synthetic spectra generated and used
for network training. A spectrum was generated at each of the four
metallicities, \met\ = $0.0$, $-0.2$, $-0.5$ and $-1.0$, yielding a
set of 155 spectra per metallicity.
The microturbulence velocity, \vmicro, was fixed at $2.0$\kms.
The temperature steps above are the finest steps in which
Kurucz's model atmospheres are available.  The gaps at high
temperature represent unstable model atmospheres which blow apart due
to high radiative fluxes (Kurucz 1992).
%% \nocite{kurucz_92a}
%% MN macro seems to not like citations in figure and table captions
\sp\ takes about one hour (on a {\sc SUN} Sparc 10) to calculate
 a synthetic spectrum in 0.02\,\AA\ steps over the 
 range 3790\,\AA--5200\,\AA.
}
\begin{tabular}{|r||c|c|c|c|c|c|}\hline
\teff & \multicolumn{6}{c|}{$\log g$} \\ \cline{2-7}
      & 2.0 & 2.5 & 3.0 & 3.5 & 4.0 & 4.5 \\ \hline
30000 &     &     &     &  x  &     &     \\
25000 &     &     &  x  &  x  &     &     \\ 
22000 &     &     &     &     &  x  &  x  \\ 
20000 &     &     &  x  &  x  &  x  &  x  \\ 
18000 &     &     &     &     &  x  &  x  \\ 
16000 &     &     &     &     &  x  &  x  \\ 
15000 &     &  x  &  x  &  x  &     &     \\ 
14000 &  x  &  x  &  x  &  x  &  x  &  x  \\ 
13000 &     &  x  &  x  &  x  &  x  &  x  \\ 
12000 &     &  x  &  x  &  x  &  x  &  x  \\ 
11000 &     &  x  &  x  &  x  &  x  &  x  \\ 
10000 &  x  &  x  &  x  &  x  &  x  &  x  \\ 
9750  &     &     &     &     &  x  &  x  \\
9500  &  x  &  x  &  x  &  x  &  x  &  x  \\
9250  &     &     &     &     &  x  &  x  \\ 
9000  &  x  &  x  &  x  &  x  &  x  &  x  \\ 
8750  &     &     &     &     &  x  &  x  \\ 
8500  &  x  &  x  &  x  &  x  &  x  &  x  \\ 
8250  &     &     &     &     &  x  &  x  \\ 
8000  &  x  &  x  &  x  &  x  &  x  &  x  \\ 
7750  &     &     &     &     &  x  &  x  \\ 
7500  &  x  &  x  &  x  &  x  &  x  &  x  \\ 
7250  &     &     &     &     &  x  &  x  \\ 
7000  &  x  &  x  &  x  &  x  &  x  &  x  \\ 
6750  &     &     &     &     &  x  &  x  \\ 
6500  &  x  &  x  &  x  &  x  &  x  &  x  \\
6250  &     &     &     &     &  x  &  x  \\ 
6000  &  x  &  x  &  x  &  x  &  x  &  x  \\ 
5750  &  x  &  x  &  x  &  x  &  x  &  x  \\ 
5500  &  x  &  x  &  x  &  x  &  x  &  x  \\
5250  &  x  &  x  &  x  &  x  &  x  &  x  \\ 
5000  &  x  &  x  &  x  &  x  &  x  &  x  \\ 
4750  &  x  &  x  &  x  &  x  &  x  &  x  \\
4500  &  x  &  x  &  x  &  x  &  x  &  x  \\ 
4250  &  x  &  x  &  x  &  x  &  x  &  x  \\ 
4000  &  x  &  x  &  x  &  x  &  x  &  x  \\ \hline
\end{tabular}
\label{synspec}
\end{center}
\end{table}
Table~\ref{synspec} lists the spectra which we generated.
The \logg\ = 4.0 and \logg\ = 4.5 models are dwarfs and the remainder
correspond to giants or subgiants. 
The spectra in Table~\ref{synspec} were calculated for
each of four metallicities, \met\ = $0.0$, $-0.2$, $-0.5$, $-1.0$, giving
a total of 155 spectra at each metallicity. A microturbulence velocity of
\vmicro\ = $2.0$\kms\ was used throughout.
\sp\ creates good quality spectra 
over a fairly wide range of spectral types, but 
it does not produce very reliable spectra for
very early- or very late-type stars.  The
generation of accurate spectra at the hot end is inhibited primarily by
the assumption of LTE, as NLTE effects are important in O-type stars.
This should not have any significant effect on the calibrations of the
MHD spectra, as the earliest type stars present are B2. At the other
end of the temperature scale, spectra cooler than about 4250\,K
(approximately K7 and later for dwarfs, and K3 and later for giants)
are not accurately synthesized on account of the absence of many
important molecules (e.g.\ H$_2$O) from Kurucz's models, and 
on account of the fact
that \sp\ does not include TiO in its opacity calculations (although
it does include MgH, C$_2$, NH, CH, CN and SiH) 
(R. Gray, private communication, 1996). The
quality of the spectra will also start to decrease for stars later
than about G2 (solar), because in these stars a significant part of
the line formation occurs in the chromosphere, which is not modelled by
\sp.

\subsection{Processing}\label{prepro}

\begin{figure}
\centerline{
\psfig{figure=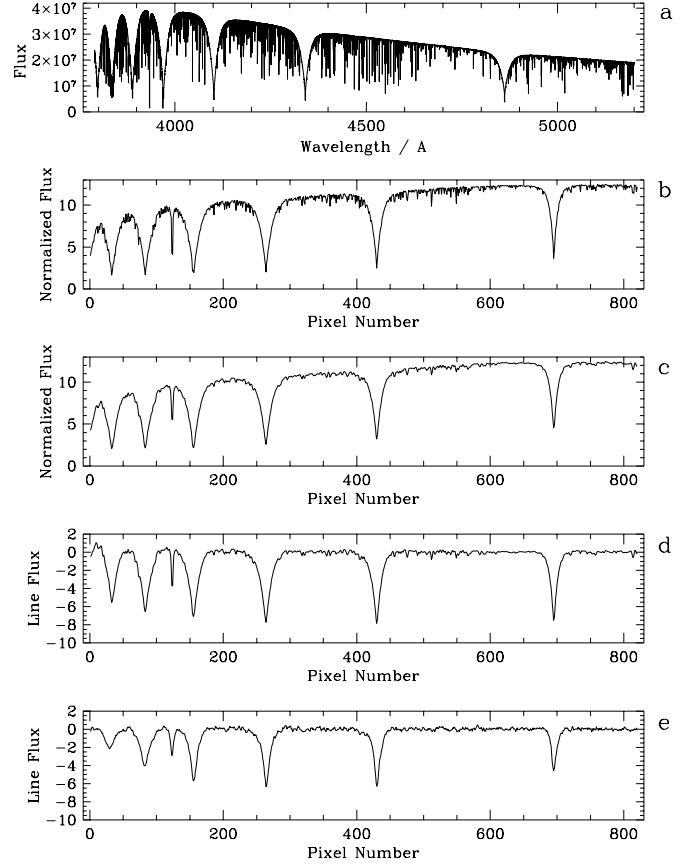,angle=0,width=0.5\textwidth}
}
\caption{Processing a synthetic spectrum. (a) A synthetic spectrum as calculated
by \sp, with parameters \met\ = 0.0, \logg\ = 4.0, \teff\ = 9000,
shown sampled at 0.1\,\AA\ on a linear wavelength scale. The flux is in
units of $erg\,s^{-1}\,cm^{-2}\,\AA^{-1}$. (b) The spectrum is
re-binned to the dispersion of the MHD spectra and area normalized.
(c) The spectrum is blurred with a $\sigma = 0.88\,pix$ ($=3''$ FWHM
`seeing') Gaussian filter.  (d) The continuum is removed.
(e) A bright (mag = 8.1) MHD spectrum (HD 23164)
of spectral class A1 V, which has similar physical parameters to the
synthetic spectrum shown in (d).}
\label{procspec}
\end{figure}

The synthetic spectrum generated by \sp\ is an energy flux spectrum
evaluated at 0.02\,\AA\ intervals.  If a neural network is to be
trained on a set of synthetic spectra and used to get physical
parameters for real observed spectra, then the observed and synthetic
spectra must be processed into a homogenous form.  Specifically, they
must have common wavelength and flux scales. Because we are free to
generate the synthetic spectra at high resolution and infinite S/N,
we chose to process these into the format of the MHD spectra, rather
than vice versa. The processing steps are summarized in
Figure~\ref{procspec}.

When non-linear processing operations are involved, the order of the
operations in this processing is relevant. It should correspond to the
effective processing performed by the telescope, disperser and
detector in obtaining a real spectrum.  The first stage of processing
the synthetic spectra was therefore to re-bin them (conserving flux in
the process) to the dispersion of the MHD spectra.  This required the
wavelength calibration of the MHD spectra. We obtained this from the
average spectrum of the MHD data set to give a calibration which
averaged out any slight wavelength misalignments incurred when
reducing the objective prism spectra.  A third-order fit to 16
identified lines in the spectrum gave an RMS wavelength--pixel
calibration error of $< 0.4$\,\AA.  As the synthetic spectra are
over-sampled relative to the real spectra by a factor of at least 50
(= 1.05\,\AA\,$pix^{-1}$/0.02\,\AA), we were justified in linearly
interpolating the synthetic spectra across their 0.02\,\AA-wide bins
when re-binning them to the MHD flux bins.  This re-binning yields a
spectrum such as that shown in Figure~\ref{procspec}b. Note that the
apparent number and strength of the metal lines below some equivalent
width is greatly reduced.

The next step was to `blur' the spectra to simulate atmospheric
turbulence (`seeing') and telescope tracking errors. 
This
blurring was achieved by convolving a Gaussian (truncated at $\pm
3\sigma$) with the spectrum. This Gaussian was normalized to conserve
flux. A smooth blurring (and flux conservation) at the end of the
spectrum was obtained by reflecting the spectrum about its ends prior
to the convolution.  We experimented with a number of different values
of the $\sigma$ parameter to obtain a best match between the
appearance of a blurred synthetic spectrum and an MHD spectrum of
similar physical properties (as inferred from the MK classification
and visual inspection).  Given that each bin in an MHD spectrum
corresponds to $1.45''$, a FWHM ($2.35\sigma$) seeing of $\theta$
arcseconds corresponds to $\sigma = 0.29\times \theta\,pix^{-1}$
arcseconds.  We found that the optimal blurring was about $3''$
($0.88$ pixels), which seems reasonable given the relatively poor
tracking ability of the Michigan Curtis Schmidt telescope.

The final stage of preprocessing was continuum removal.  This was done
in exactly the same way as for the MHD spectra: a spectral continuum
is produced by median and then boxcar filtering the spectrum. This
continuum is subtracted from the original spectrum to yield a
rectified (line-only) spectrum.  To improve the continuum fit in the
region of broad lines (most importantly H lines, the CN band and Ca II
H\&K lines), these regions were masked off prior to the median
filtering.  Further details are given in Bailer-Jones et~al.\
(1997b)\nocite{bailerjones_97b}.

The final spectrum is shown in Figure~\ref{procspec}d above a real
spectrum of similar physical parameters. The removal of the
continuum from the 
synthetic spectra gives a good line-only spectrum everywhere apart
 from at the very blue end.  This was not apparent in the continuum
removal of the MHD spectra as the plate QE is very low at the blue end
of the spectrum.  To improve the match between the spectra,
we chose to exclude the first ten flux bins from all synthetic and MHD
spectra when using them together in a neural network. The final
spectra therefore consist of 810 flux bins.

\section{Neural Network Calibration}

\subsection{\mbox{\boldmath $T_{\rm{eff}}$} Calibration}\label{tcalsec}

To determine effective temperature, \teff, for the MHD spectra, a
neural network was set-up with a single continuous output to represent
\teff.  The type of neural network used had an output range confined
to 0--1. In order to fit the 4000\,K to 30~000\,K temperature range of
the synthetic models comfortably into this range, we used the
transformation $T = (\log$ \teff$- 3)/2$, where $T$ is the network
target. A more straightforward implementation would be to use a
network with linear outputs which has an unbounded output range,
although we would probably still want to use log~\teff\ as the target
in order to reduce the dynamic range. The exact transformation used
should not matter.

Following on from the work of Bailer-Jones et~al.\
(1997a)\nocite{bailerjones_97a}, we used an 810:5:5:1 neural network
architecture for this calibration problem. This refers to a neural
network with 810 inputs (the number of flux values in the spectrum), 5 nodes
in each of two hidden layers (Figure~\ref{typnetarch.p3} shows a network
with only a single hidden layer) and a single output. 

\begin{figure}
\centerline{
\psfig{figure=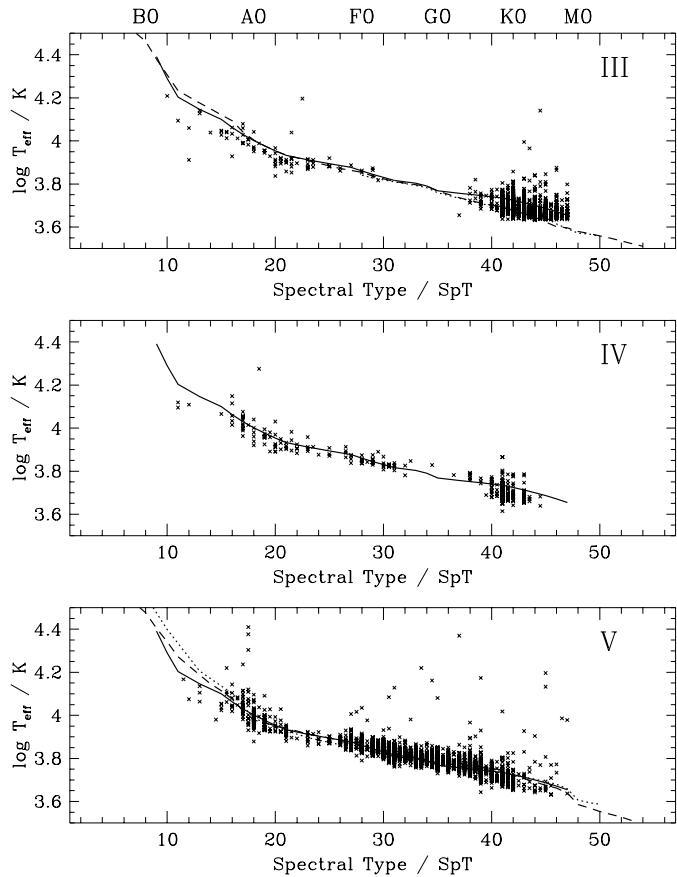,angle=0,width=0.5\textwidth}
}
\caption{\teff--SpT relationship for the MHD spectra from a committee
of neural networks trained on
\met\ = $-0.2$ synthetic spectra. The relationships for
luminosity classes III, IV and V are shown separately, although the
values of \teff\ for these were obtained from the same committee.  The
lines show calibrations from the literature for comparison. The solid
line is the calibration of Gray and Corbally (1994); 
%% \nocite{gray_94b}
the
dashed line is that of Schmidt-Kaler (1982); 
%% \nocite{schmidtkaler_82a}
the
dotted line is that of Gray (1992). 
%% \nocite{gray_92a}
Gray and Corbally only
give a calibration of dwarf (V) stars, but this calibration line has
been included in the top two plots to ease comparison between the
plots of the neural network calibrations.
Schmidt-Kaler and Gray both give separate
calibrations for giants (III). None give calibrations for the
subgiants (IV).
}
\label{tsyn03}
\end{figure}

Rather than using a single neural network to give predictions of
\teff, we used a committee of ten neural networks differing only in
their initial random weights. The ouputs
given by each member of the committee are averaged to give a single
prediction which is generally more reliable than that given by any
single network.  Four training data sets were constructed, one for
each of the four metallicities at which synthetic spectra were
generated: \met\ = $0.0$, $-0.2$, $-0.5$ and $-1.0$. Thus each
training set consisted of 155 spectra of the same metallicity and with
values of \teff\ and \logg\ shown in Table~\ref{synspec}.
A separate committee was trained on each of the four data sets.  
Once trained,
the committees were used to evaluate \teff\ for all of the MHD
spectra.  Figure~\ref{tsyn03} shows the \teff-SpT calibration produced
by the committee trained on the \met\ = $-0.2$ spectra, over-plotted
by calibrations from the literature.

The MK spectral type classifications along the $x$-axis are the
catalogue classifications of the MHD spectra as given by Houk (1975,
1978, 1982,
1988)\nocite{houk_75a}\nocite{houk_78a}\nocite{houk_82a}\nocite{houk_88a},
converted into codes on a 1--57 numerical scale as shown in
Table~\ref{caltab}.  These classifications are all either integral or
half-integral spectral subtypes, so the data are discrete along the
$x$-axis.  Because the synthetic spectra are only valid for
temperatures above about 4250\,K, we would not expect the neural
networks to give reliable calibrations for spectra later than about
K5, and we have omitted all spectra later than K5 from these
calibration plots.

Apart from a handful of outliers, the majority of the 2732
dwarfs in the bottom panel of Figure~\ref{tsyn03} show a reasonably
`tight' correlation between MK spectral type and
\teff. Their distribution also agrees quite well with the
published calibrations.  The giant stars also show a reasonable
\teff--SpT correlation,  although it deteriorates towards cooler
stars, most probably due to the lower quality and accuracy of the
synthetic spectra in this region.  We have not discovered any
\teff--SpT calibrations in the literature with which to compare the
subgiant calibrations, but our data shows a reasonably tight correlation
between \teff\ and spectral type.

\begin{figure}
\centerline{
\psfig{figure=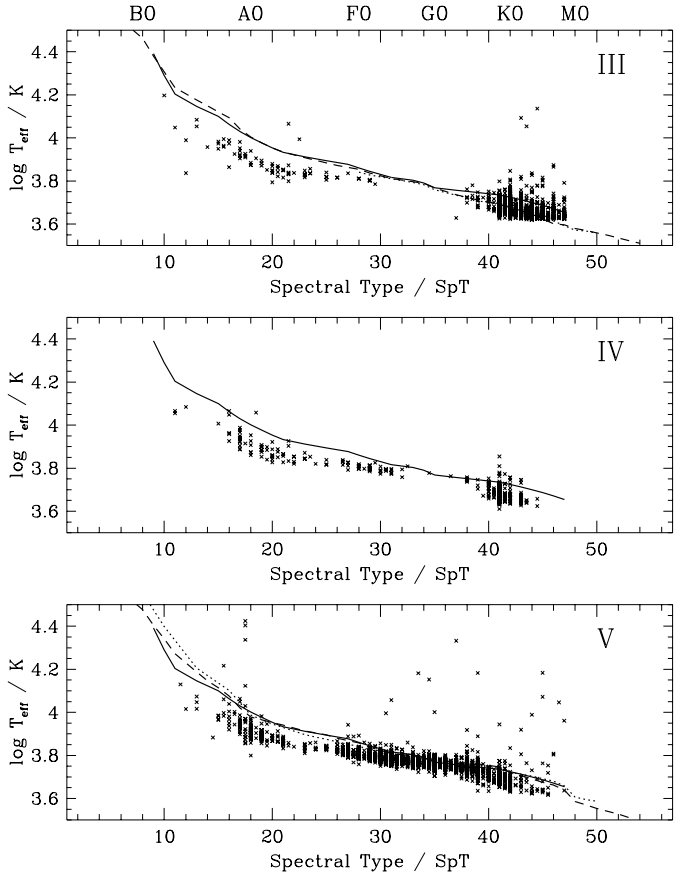,angle=0,width=0.5\textwidth}
}
\caption{\teff--SpT relationship for the MHD spectra from a committee
of neural networks trained on \met\ = $-1.0$ synthetic spectra.  See
caption to Figure~\ref{tsyn03} for further details.}
\label{tsyn05}
\end{figure}

As a comparison, Figure~\ref{tsyn05} shows a similar plot to
Figure~\ref{tsyn03} for a committee of ten networks trained on the
\met\ = $-1.0$ spectra.
Here we see a slightly tighter correlation for the dwarfs, but the
agreement with the published calibrations is not as good. The
correlations for networks trained on \met\ = $-0.5$ and \met\ = $0.0$
spectra were also quite tight, but they did not agree with the literature
calibrations for both giants and dwarfs as well as \met\ = $-0.2$.
Metallicity effects on the calibration are discussed in
section~\ref{meteff} below.

The data in Figure~\ref{tsyn03} can be used to give a statistical
calibration between \teff\ and spectral type. The most suitable way of
achieving this is by forming the frequency distribution, \td, of
\teff\ for each spectral type separately, where $s$ indicates that $D$
is different for each spectral type. With a sufficiently large amount
of data, \td\ would be approximately Gaussian. The median of \td,
\mtf, is a robust measure of the \teff\ calibration for this spectral
type.  An appropriate measure of the spread of \td\ is \sig68.  This
is the value which confines 68\% of the data, and is equal to
$1\sigma$ if the distribution is Gaussian.  For some spectral types
there were only a few spectra; in these cases it was necessary to
linearly interpolate \td\ in order to yield an accurate measure of
\sig68.

\begin{figure*}
\begin{minipage}{10cm}
\centerline{
\psfig{figure=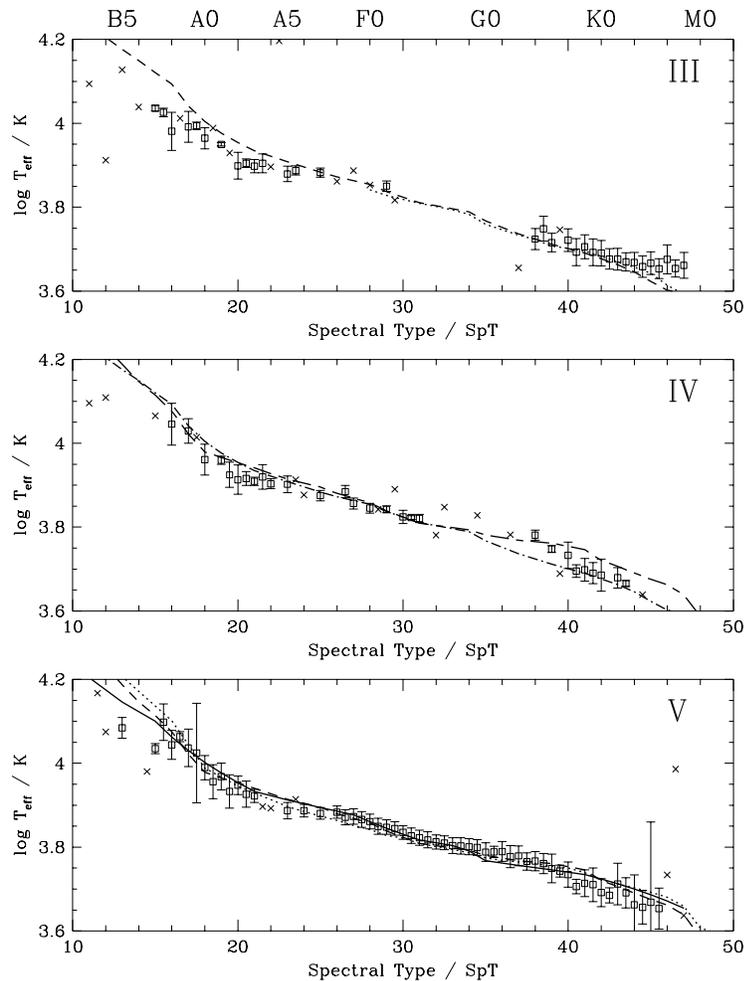,angle=0,width=1.0\textwidth}
}
\caption{\teff--SpT calibration of data shown in Figure~\ref{tsyn03}
(\met\ = $-0.2$).  The squares correspond to the median, \mtf, of the
frequency distribution of \teff\ for each spectral type. The error
bars are the \sig68\ errors, i.e.\ those bounds which include 68\% of
the data. Those spectral types for which there were only one or two
points in the distribution are marked with a cross.  These points are
not very reliable and as the error cannot be reliably determined, it is
not shown.  The overplotted lines are the calibrations of Gray and
Corbally (1994) (solid line),
%% \nocite{gray_94b} 
Schmidt-Kaler (1982)(dashed line)
%% \nocite{schmidtkaler_82a} 
and Gray (1992) (dotted line).  
%% \nocite{gray_92a}
We found no published calibration for
the subgiants (IV) and instead have overplotted the middle figure with
the giant (dot-dash line) and dwarf (dash line) calibrations of
Schmidt-Kaler. Note that the axis scales are slightly different here
 from those used in Figure~\ref{tsyn03}.
}
\label{calsp03}
\end{minipage}
\end{figure*}
Figure~\ref{calsp03} plots these calibrations and their \sig68\
errors. It shows a very good agreement with the published
calibrations, particularly for the dwarf stars.  At the ends of the
spectral type scale the calculated errors become significantly larger.
This could be partly on account of the increased difficulty of the
neural network to interpolate its mapping function at the ends of the
output range, as the interpolation is less well constrained in these
regions.  The larger errors at the cool end are more likely, however,
to be attributable to the lower quality of the stellar models and
synthetic spectra at low temperatures.  The larger errors at high
temperatures, SpT~$\ltsim 20$, are probably an indication of a poorer
determination of the \teff--SpT relationship which in turn is the
result of there being fewer spectra in this region.
Table~\ref{caltab} tabulates the calibrations and errors in
Figure~\ref{calsp03} and shows the number of spectra, $N$, of each
spectral type used to determine \mtf\ and \sig68.  
Note that the
literature calibrations will also have some degree of uncertainty,
although as the respective authors do not provide details of this it
is difficult to perform rigorous statistical tests of the level of
agreement between our calibrations and the literature ones.
\begin{table*}
\begin{minipage}{14cm}
\begin{center}
\caption{\teff--MK spectral type calibrations from a committee of
neural networks trained on \met\ = $-0.2$ spectra. These values are
the tabulations of those shown in Figure~\ref{calsp03}; \mtf\ is the
median calibration temperature and \sig68\ is its error bar shown in
Figure~\ref{calsp03}. $N$ is the number of stars of each spectral
type, and hence the number of points used to define the distribution
 from which \mtf\ and \sig68\ are determined. Note that \sig68\
measures the spread of this distribution and is {\it not} the standard
error on the median, which is $\sigma_{68}\sqrt{\pi/2N}$.}
\begin{tabular}{|r|r||rcr|r||rcr|r||rcr|r|}\hline
\multicolumn{2}{c||}{} & \multicolumn{4}{c||}{Dwarfs (V)} & \multicolumn{4}{c||}{Subgiants (IV)} & \multicolumn{4}{c|}{Giants (III)} \\ \hline
\multicolumn{2}{|c||}{Spectral} & \multicolumn{2}{}{} & & & \multicolumn{2}{}{} & & & \multicolumn{2}{}{} & & \\ 
\multicolumn{2}{|c||}{Type}     & \multicolumn{2}{c}{\mtf\,/\,K} & \sig68 & $N$ & \multicolumn{2}{c}{\mtf\,/\,K} & \sig68 & $N$ &  \multicolumn{2}{c}{\mtf\,/\,K} & \sig68 & $N$ \\ \hline
13 &    B5 &  12145 & $\pm$ &  685 &   3        &&&&&&&& \\             
14 &    B6 &  &&&&&&&&&&& \\                    
15 &    B7 &  10842 & $\pm$ &  301 &   4        &&&&                            & 10868 & $\pm$ &  173 &  3 \\
16 &    B8 &  11061 & $\pm$ &  884 &   7        & 11106 & $\pm$ & 1276  &  5    & 9568  & $\pm$ & 1002 &  3 \\
17 &    B9 &  10877 & $\pm$ & 1118 &  21        & 10695 & $\pm$ &  723  & 13    & 9806  & $\pm$ &  836 &  6 \\
18 &    A0 &   9762 & $\pm$ &  651 &  51        & 9139  & $\pm$ &  779  &  6    & 9211  & $\pm$ &  538 &  4 \\
19 &    A1 &   9298 & $\pm$ &  678 &  26        & 9109  & $\pm$ &  206  &  7    & 8907  & $\pm$ &   84 &  3 \\
20 &    A2 &   8862 & $\pm$ &  454 &   7        & 8190  & $\pm$ &  662  &  8    & 7919  & $\pm$ &  580 &  7 \\
21 &    A3 &   8364 & $\pm$ &  309 &  16        & 8110  & $\pm$ &  182  &  4    & 7903  & $\pm$ &  285 &  4 \\ 
22 &    A4 &        &       &      &            & 8006  & $\pm$ &  213  &  4    &&&& \\
23 &    A5 &   7712 & $\pm$ &  347 &   9        & 7982  & $\pm$ &  360  &  4    & 7573  & $\pm$ &  323 &  7 \\
24 &    A6 &   7711 & $\pm$ &  266 &   6        & &&&&&&& \\    
25 &    A7 &   7591 & $\pm$ &  242 &   8        & 7503  & $\pm$ &  219  &  3    & 7628  & $\pm$ &  192 &  4 \\
26 &    A8 &   7645 & $\pm$ &  259 &  12        & &&&&&&& \\    
27 &    A9 &   7466 & $\pm$ &  311 &  76        & 7188  & $\pm$ &  217  &  4 &&&& \\    
28 &    F0 &   7261 & $\pm$ &  295 & 122        & 7003  & $\pm$ &  187  &  6 &&&& \\    
29 &    F2 &   7040 & $\pm$ &  281 & 135        & 6956  & $\pm$ &  132  &  9    & 7081  & $\pm$ &  207 &  3 \\
30 &    F3 &   6838 & $\pm$ &  257 & 265        & 6672  & $\pm$ &  244  &  5 &&&& \\    
31 &    F5 &   6643 & $\pm$ &  280 & 269        & 6618  & $\pm$ &  147  &  5 &&&& \\    
32 &    F6 &   6492 & $\pm$ &  254 & 146        & &&&&&&& \\    
33 &    F7 &   6364 & $\pm$ &  275 & 170        & &&&&&&& \\    
34 &    F8 &   6324 & $\pm$ &  275 &  65        & &&&&&&& \\    
35 &    G0 &   6141 & $\pm$ &  316 & 145        & &&&&&&& \\    
36 &    G1 &   6161 & $\pm$ &  333 &  83        & &&&&&&& \\    
37 &    G2 &   6015 & $\pm$ &  328 &  70        & &&&&&&& \\    
38 &    G3 &   5843 & $\pm$ &  307 & 157        & 6041  & $\pm$ &  166  &  6    & 5297  & $\pm$ &  309 &  4 \\
39 &    G5 &   5613 & $\pm$ &  438 & 116        & 5598  & $\pm$ &  105  &  3    & 5193  & $\pm$ &  265 & 10 \\
40 &    G6 &   5429 & $\pm$ &  373 &  50        & 5406  & $\pm$ &  389  & 17    & 5263  & $\pm$ &  329 & 15 \\
41 &    G8 &   5171 & $\pm$ &  381 &  59        & 4992  & $\pm$ &  311  &132    & 5076  & $\pm$ &  332 &154 \\
42 &    K0 &   4914 & $\pm$ &  378 &  31        & 4847  & $\pm$ &  426  & 20    & 4903  & $\pm$ &  344 &387 \\
43 &    K1 &   5149 & $\pm$ &  586 &  11        & 4780  & $\pm$ &  268  & 23    & 4744  & $\pm$ &  282 &285 \\
44 &    K2 &   4601 & $\pm$ &  750 &   7        &       &       &       &       & 4658  & $\pm$ &  254 &198 \\
45 &    K3 &   4662 & $\pm$ & 2057 &   9        &       &       &       &       & 4642  & $\pm$ &  290 &105 \\
46 &    K4 & &&&&&&&                                                            & 4736  & $\pm$ &  379 & 53 \\
47 &    K5 & &&&&&&&                                                            & 4585  & $\pm$ &  328 & 30 \\ \hline
\end{tabular}
\label{caltab}
\end{center}
\end{minipage}
\end{table*}

It is important to realise that the MK spectral type parameter is not a 
continuous variable. It is a set of discrete classes, each of which does
not correspond to a single unique temperature.
There is, therefore, a cosmic scatter of temperature about
the median calibration value, \mtf, and this intrinsic scatter will make a
significant contribution to \sig68 (other contributions are discussed
below).  The important point is that
\sig68\ is {\it not} a measure of the error with
which \mtf\ has been determined.  The error in \mtf\ itself is given
by the standard error in the median, which is smaller than \sig68\
by a factor of $0.79\sqrt{N}$ ($\equiv \sqrt{2N/\pi}$) for large $N$.  
This error for F5 dwarfs, for example,
is $280/(0.79\sqrt{269}) = 22$\,K.  Almost all of the
values of \td\ in Table~\ref{caltab} have
standard errors in the median of less than 50\,K. This is
about as small as can be meaningfully reported. In comparison,
Jones, Gimore \& Wyse (1996)\nocite{jones_96a} 
used a broad band photometric index,
$(V-I)_{c}$, to determine temperature. 
They showed that an error in this index of
$\pm 0\!\!\stackrel{m}{.}\!\!05$ gives rise to a temperature error of
120--240\,K (depending upon the spectral type). An error of $\pm
0\!\!\stackrel{m}{.}\!\!02$ (which is about the limit of the photometry)
would yield errors of 47--97\,K.

Cosmic scatter is not the only factor contributing to \sig68.  Another
source of error is the simplifying assumptions used in generating the
atmospheric models and synthetic spectra, notably the assumption of
LTE and the absence of certain atoms and molecules.  Additionally, the
synthetic spectra do not display variances in other stellar properties
which will show cosmic variance, such as the relative abundances of
atomic and molecular species or \vmicro.  These will produce
variations in the MHD spectra which are `unexpected' by a network
trained on the synthetic spectra.  Another source of error is likely
to derive from the neglect of line formation mechanisms in the
chromosphere.

Whilst we made efforts to process the synthetic spectra into the same
format as the MHD spectra, they still differ in the possibly important
respect that the synthetic spectra are on a linear flux scale, whereas
the MHD spectra are on a non-linear flux scale.  This difference could
be removed by flux calibrating the photographic plates (e.g.\ by
identifying spectrophotometric standards on the plates).  A final
problem could be that the synthetic spectra have an unrealistically
high S/N. This could lead the networks to lock onto certain weak
features in the synthetic spectra that cannot be used as a source of
class discrimination in the lower S/N MHD spectra. One way around this
would be to add random noise to the synthetic spectra and use each
synthetic spectrum several times in the training set with different
random noise added in each case.  
Nonetheless, the narrow distribution
of residuals about the median calibrations demonstrate
that these potential problems have only a small effect on our
technique and on the resulting calibration shown in Figure~\ref{calsp03}
and Table~\ref{caltab}.

\subsection{Metallicity Effects}\label{meteff}

It is apparent from Figures~\ref{tsyn03} and~\ref{tsyn05} that, when
the metallicity of the synthetic spectra is decreased, the neural
networks give {\it systematically} lower effective temperatures for
stars of a given spectral type. The calibration results from training
neural networks on synthetic spectra of \met\ = $0.0$ and of \met\ =
$-0.5$ confirm this trend. 
An explanation of this will be discussed shortly.
This is an important result because it demonstrates that the
network is sensitive to metallicity, and that in principle this
sensitivity can be exploited to give metallicity calibrations for the
MHD spectra. How this could be done will be discussed later.

The best agreement between our neural network calibrations and the
literature calibrations was obtained when the networks were trained on
\met\ = $-0.2$ spectra. The literature calibrations are of course based on
the same original MK standard spectra; hence the metallicity at which an
agreement is achieved tells us something about the mean metallicity of
the MHD stars and the MK system in general.  That this comes out to
\met\ = $-0.2$ is not surprising, as it is well known that 
\met\ = $-0.2$ for the local disk (see, for example, 
Wyse \& Gilmore, 1995\nocite{wyse_95a}, and references therein to
earlier work). If the distances to the MHD stars
are calculated and 
plotted as a histogram, the giants and dwarfs show a peak in the
distribution at $\approx 400\,pc$ and $\approx 100\,pc$ respectively
(Bailer-Jones 1996).
For many of the stars, which are at high Galactic latitudes, 
these distances are approximately distances out
of the Galactic plane.  The MHD stars are therefore primarily Galactic thin
disk stars. The thin disk has exponential scale heights of $\approx
100\,pc$ and $\approx 300\,pc$ for the young and old stars respectively
(Gilmore \& Wyse 1985)\nocite{gilmore_85a}.  Fitting a Gaussian to the
distribution of stars plotted against [Fe/H], Gilmore and Wyse
(1985)\nocite{gilmore_85a} reported a mean abundance
$\overline{\rm{[Fe/H]}} = 0.0$ and $\sigma_{\rm{[Fe/H]}} = 0.15$ for the young
thin disk, and $\overline{\rm{[Fe/H]}} = -0.3$ and $\sigma_{\rm{[Fe/H]}} = 0.2$
for the old thin disk.  (As \sp\ uses standard relative abundances,
\met\ $\approx$ [Fe/H] for disk stars.)  It is therefore plausible
that the MHD spectra have an average metallicity of $-0.2$. It is
certainly unlikely to be as high as 0.0 or as low as $-0.5$ or $-1.0$.
By comparison, the metallicity fit for the Galactic thick disk (scale
height~$\gtsim 1\,kpc$) is $\overline{\rm{[Fe/H]}} = -0.6$ and
$\sigma_{\rm{[Fe/H]}} = 0.3$, and for the Galactic extreme spheroid (scale
height~$\gtsim 4\,kpc$) it is $\overline{\rm{[Fe/H]}} = -1.5$ and
$\sigma_{\rm{[Fe/H]}} = 0.5$.  

\begin{figure}
\centerline{
\psfig{figure=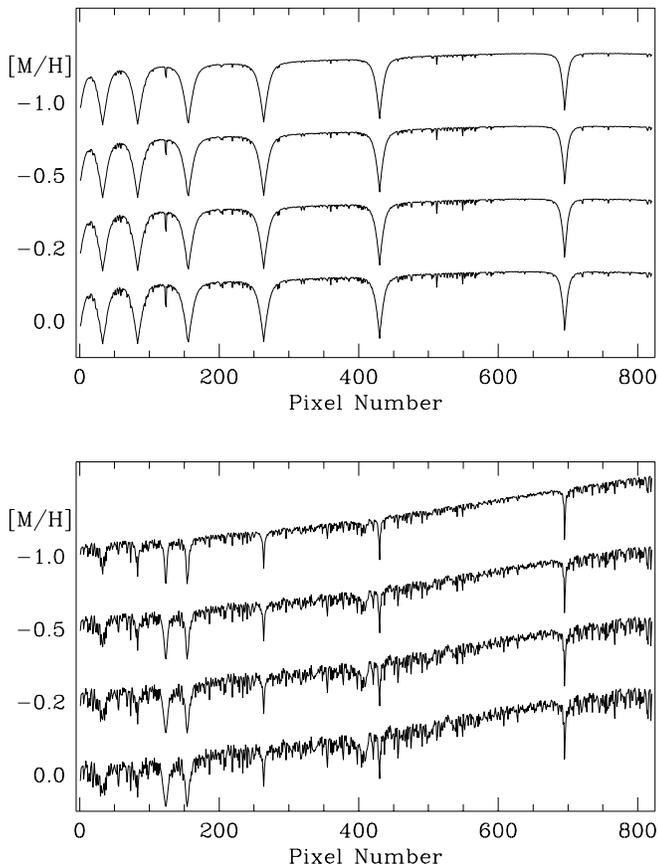,angle=0,width=0.5\textwidth}
}
\caption{Metallicity effects in the synthetic spectra. The top
four spectra are \teff\ = 10~000 K, \logg\ = 4.0. The bottom four
are  \teff\ = 6250 K, \logg\ = 4.0. In each case metallicity increases
down the box, with the value of \met\ shown on the left. The spectra
have been re-binned to the dispersion of the MHD spectra,
but have not been blurred.} 
\label{metal}
\end{figure}

Why do the \teff\ calibrations change with metallicity?  The strength
of the metal lines in a synthetic spectrum increases with increasing
metallicity, as can be seen from Figure~\ref{metal}.
However, at a fixed metallicity, the apparent number and strength of
the metal lines also generally increases as \teff\ decreases.  Suppose
a network is trained on metal rich stars and gives the \teff\ of a
certain MHD star as $T$. A neural network trained on metal weak stars
would assign this MHD star a temperature {\it lower} than $T$, because this MHD
spectrum is closer in appearance to a cooler star in the set of
metal weak spectra.  In other words,
because the neural networks are trained on synthetic spectra of a
single metallicity, they are not entirely able to distinguish
metallicity features from temperature ones when attempting to classify
MHD spectra.  We tried to solve this problem by training a network on
a set of spectra of all four metallicities (a total of 620 spectra),
with the hope that the networks could marginalize over metallicity
effects. But this gave even poorer calibrations, both in terms of
poorer agreement with the published calibration and a poorer
\teff--SpT correlation.

This confusion of temperature effects with metallicity ones is not
just a problem for neural networks. There is a real
correlation between metallicity and temperature.  At lower
temperatures there is less excitation and ionization of the metals,
and hence more metal lines in the spectrum.  Although the \met\ value
of the MHD star is constant, it is difficult for the neural network to
isolate those features which vary only with metallicity and not with
temperature. Thus simply training a neural network on whole spectra with a
range of metallicities is unlikely to help, because, rather than
permitting the network to `ignore' metallicity effects, this is more
likely to mislead it.

It is clearly essential that metallicity effects are considered when
attempting to obtain \teff\ calibrations for spectra.  If the
metallicity of the observed spectrum could be determined first, then
it could be used as an extra input to a network along with the
spectrum. If this network were trained on synthetic spectra with a
range of metallicities, then we could imagine that the additional
metallicity information would permit an appropriate \teff\
determination. A more suitable alternative would be to have a series
of networks, each trained on spectra of a single metallicity, and then
to use the network with the appropriate metallicity to evaluate \teff:
This is more or less what we did earlier by selecting that metallicity
which gave the best agreement with existing \teff--MK calibrations.
The drawback with these approaches is that they require the
metallicity to be known in advance.  Jones, Wyse \& Gilmore
(1995)\nocite{jones_95a} and Jones~et~al.\ (1996)\nocite{jones_96a}
describe a method of determining metallicities from relatively low S/N
ratio (10--20) spectra, although at a higher resolution than the MHD
spectra.  They use spectroscopic indices calibrated by synthetic
spectra to measure [Fe/H] and avoid confusion with temperature by
using an independent photometric measure of \teff.

Despite the confusion between temperature and metallicity,
Figures~\ref{tsyn03} and~\ref{tsyn05} nonetheless show that
metallicity produces a {\it systematic} rather than a {\it random}
effect on the \teff--SpT correlation. In principal, therefore, this
metallicity signature can be exploited by the neural networks to
obtain metallicities for the MHD stars.  The question is how to do it.
One solution may to be to determine \met\ and \teff\ simultaneously,
e.g.\ by using a neural network with two outputs.  A direct approach
to determining metallicity independently of temperature may work by
identifying only those spectral regions which are most sensitive to
metallicity and least sensitive to temperature, such as certain iron
lines in late-type stars (Keenan\& McNeil 1976)\nocite{keenan_76a}.
With only these regions as inputs to a neural network, the `noise'
caused by temperature variations is greatly reduced, making it more
likely that the neural network could recognise metallicity features.
This is an area for future work.

\section{Summary}

We have shown that neural networks trained on synthetic spectra
provide low-error predictions for the effective temperature, \teff, of
a star based on the star's optical spectrum. By applying these neural
networks to spectra with exisiting MK classifications, we have
obtained a calibration between the MK spectral type parameter and
\teff. This calibration shows a good agreement with a number of
calibrations from the literature.  The calibration was obtained
statistically from a number of optical spectra at each spectral type
with a resolution of $\approx$ 2\,\AA\,$pix^{-1}$.  The precision of
this calibration is largely limited by the cosmic scatter in
temperature for a given MK class and by limitations of the stellar
models. It has also been shown that metallicity effects have to be
considered when trying to determine \teff. Further work is required
before neural networks can be used to accurately quantify
metallicities. In particular, further processing of the synthetic
spectra into the format of the MHD spectra (e.g.\ by the addition of
noise and calibration of the flux scale) may be required.  Our work
nonetheless demonstrates that our networks are sensitive to
metallicity.

\section*{Acknowledgments}

We would like to thank Nancy Houk for kindly loaning us her plate
material. We are grateful to Richard Gray for the use of his spectral
synthesis program and his effort in converting the code to operate on
our computer system.  We would also like to thank Robert Kurucz for
the use of his model atmospheres.

%%%%%%%%%%%%%%%%%%%%%%%%%% BIBLIOGRAPHY %%%%%%%%%%%%%%%%%%%%%%%%%%%%%%

\end{document}